\documentclass[preprint,5p,times,twocolumn]{elsarticle}

\usepackage{amsmath}
\usepackage{amssymb}
\usepackage{graphicx}
\usepackage{booktabs}
\usepackage{color}
\usepackage{hyperref}
\usepackage{xspace}
\usepackage{braket}
\usepackage{dsfont}
\usepackage{lipsum}
\usepackage{slashed}

\journal{Physics Letters B}

\begin{document}

\begin{frontmatter}

\title{An improved Bell-CHSH observable for gauge boson pairs}
  
\author[first,second]{Rados\l{}aw Grabarczyk}\ead{radoslaw.grabarczyk@physics.ox.ac.uk}
\affiliation[first]{organization={Department of Physics, University of Oxford},
            addressline={Keble Road}, 
            city={Oxford},
            postcode={OX1 3RH}, 
            country={UK}}
\affiliation[second]{organization={Rudolf Peierls Centre for Theoretical Physics},
            addressline={Parks Road}, 
            city={Oxford},
            postcode={OX1 3PU}, 
            country={UK}}

\begin{abstract}
For particles decaying without parity violation, it is impossible to reconstruct the full spin-density matrix from the velocities of their decay products.
In this work, we consider Bell inequalities based on squares of spin operators. The corresponding Bell operators probe only the part of the density matrix that can be reconstructed from any gauge boson decay or splitting to fermions, which is accessible regardless of whether parity is violated.
We find that our new choice of Bell operator has promising properties regarding states seen at colliders. In particular, it substantially outperforms the inequality previously used for the $pp \rightarrow ZZ$ process at the LHC in terms of the phase space volume in which it is violated. We also conduct the first investigation of Bell violation in a system involving gauge bosons mediating different interactions, namely the $W$ boson and gluon in the $pp \rightarrow W^{\pm} (g\rightarrow b\bar{b})$ process.
\end{abstract} 


\end{frontmatter}
%{\small
%\tableofcontents
%}

%======================================================================
\section{Introduction}
Measuring the the joint spin-density matrix of particles produced at the particle colliders offers a unique opportunity to study quantum phenomena at high energies. 
This representation of the quantum state can be used, among other applications, to 
detect quantum entanglement~\cite{ATLASttbar,CMSttbar, Barr_review} and measure the existence of Bell-violating states~\cite{Barr_2022, Fabbrichesi_2023, Aguilar_Saavedra_2023, Ashby_Pickering_2023, Squared-spin-Bi} of pairs of extremely short-lived fundamental particles. While such measurements cannot
be used to reject all local hidden-variable models~\cite{Dreiner, NewLHVM}, they are good tests of the internal consistency of quantum mechanics in this regime.

The main idea allowing these measurements is the fact that when a particle decays weakly, we can reconstruct its full density matrix from the angular distributions of decay products~\cite{Barr_review}.
This work aims to extend measurements of Bell-violating states to decays and splittings beyond weak interactions. All other Standard Model couplings of gauge bosons to fermions are invariant under parity transformations.
Thus, the rest-frame direction of a daughter fermion's velocity provides information about the axis of the mother particle's spin post-measurement but not its spin value along that axis.
In other words, with such interactions, we can only access the parity-invariant component of the full density matrix.
Using this component of the density matrix, we can only calculate expectation values of parity-invariant operators.

Such a partially-reconstructed density matrix cannot be used to test Bell violation with most Bell operators previously considered for pairs of qutrits, as almost none of them are parity invariant, the notable exception being studied in~\cite{Caban}.
We investigate this Bell operator alongside our new proposal. Both operators generate variants of the Clauser-Horne-Shimony-Holt (CHSH) inequality~\cite{CHSH}.

We begin by defining the two emergent inequalities and providing expressions for the Bell observables in terms of rest-frame angular distributions of daughter fermions in the absence of fiducial cuts.
We then investigate Bell inequality violation in two different processes. 

The first one is electroweak $ZZ$ production at the LHC via quark fusion, a ``golden channel'' for density matrix reconstruction, thanks to its clean 4-lepton signal.
While Bell violation in this process can be studied using the Collins-Gisin-Linden-Massar-Popescu (CGLMP) inequality~\cite{CGLMP}, it has been shown that measuring violation requires strong cuts that leave $O(10)$ events in LHC Run 2+3 data, making
an accurate density matrix reconstruction infeasible~\cite{Fabbrichesi_2023}. We show that, with the same dataset, one can probe Bell violation with almost 1000 times more events using our choice of Bell operator. This makes the measurement promising, especially for the High-Luminosity LHC (HL-LHC).

The second process is the quark-fusion production of a $W$ boson and an off-shell gluon at the LHC, where the latter splits into a $b \bar{b}$ pair.
At the lowest order in EW and QCD couplings, this is the only topology that leads to the $l \nu_l b \bar{b}$ final state, and its polarized amplitudes share
many similarities with electroweak $ZZ$ production. Using our Bell operator in this process constitutes the first proposal involving a Bell-like observable
for a quantum state containing gauge bosons mediating different fundamental interactions. 
\section{The CHSH inequality revisited}

The CHSH inequality takes the form~\cite{CHSH}
\begin{equation}
\left| E(a,b) - E(a,b') + E(a',b) + E(a',b')\right| \leq 2,
\end{equation}
where $E(a,b)$ is the product of outcomes of measurements performed on two particles in consideration, given ``settings" $a, a'$ and $b, b'$. The measurements must return a value from the set $\{-1, 0, 1\}$. 
In quantum mechanics, one can rewrite the inequality in terms of a Bell operator
\begin{equation}
  \left|\text{tr}\left(\hat{\mathcal{B}} \rho\right)\right| \leq 2,
\end{equation}
where
\begin{equation}
\hat{\mathcal{B}} = \hat{A} \otimes (\hat{B} - \hat{B}') + \hat{A}' \otimes (\hat{B} + \hat{B}'),
\end{equation}
$\hat{A}, \hat{A}', \hat{B}, \hat{B}'$ are operators representing the measurements with settings $a, a', b, b'$ respectively, and $\rho$ is the joint density matrix of the pair of particles. We are free to choose the measurement operators as long as they have eigenvalues belonging to $\{-1, 0, 1\}$. 
We focus on two choices of measurements that give parity-invariant Bell operators.
\subsection{Measurement operators}
A basis is chosen in what follows such that the particle considered is moving along the $z$-axis, and the $x$-axis lies in the plane of the particle's velocity vector and the beam.

In~\cite{Caban}, a choice based on a ``Linear Polarization" measurement was considered. We will refer to this variant as the Linear-Polarizer-Bell (LP-Bell) inequality.
The measurements correspond to different angle settings of a linear polarizer $\alpha$. Explicitly, they are represented by a family of operators of the form
\begin{align}
  &\hat{O}^{\rm{LP}}_{\alpha} = \left(\cos(\alpha) \hat{S}_x + \sin(\alpha)\hat{S}_y\right)^2 - \nonumber\\&\left(-\sin(\alpha) \hat{S}_x + \cos(\alpha) \hat{S}_y\right)^2= \cos(2 \alpha) \lambda_4 + \sin(2 \alpha) \lambda_5. 
\end{align}
where $\hat{S}_i$ is a spin-1 operator along Cartesian axis $i$ and $\lambda_j$ is the $j$th Gell-Mann matrix. 
The eigenvalues of the LP-Bell measurement operators are $+1$, $0$ and $-1$; $+1$ is assigned to the measurement of $\ket{+}$ or $\ket{-}$ spin state at angle $\alpha$ to the $x$-axis, $-1$ is assigned to such states at angle $\alpha + \pi/2$ to the $x$-axis and 0 is assigned to the longitudinal polarization.
It was shown that the LP-Bell inequality is maximally violated for a scalar state of relativistic vector bosons as long as their rest masses are negligible compared to their energies. Longitudinal polarizations, being assigned the value 0, reduce the effect at lower energies.
Noting that the statistical error on measurements of joint spin states from the distribution of decay products is large~\cite{theo, Barr_review, Fabbrichesi_2023}, we would like to construct an inequality that is expected to be violated in a kinematic regime where the cross sections are enhanced. In such regimes, vector bosons are often longitudinally polarized.

In this work, we consider measurement operators of the form 
\begin{align}
  \hat{O}^{\text{SS}}_{\hat{n}} =& 2 \left(\hat{n} \cdot \hat{\vec{S}}\right)^2 - \mathds{1}_3.
\end{align}
We will refer to this variant as the Spin-Squared-Bell (SS-Bell) inequality.
The eigenvalues of these operators are $+1$ and $-1$; $+1$ is assigned to the $\ket{+}$ or $\ket{-}$ spin state along $\hat{n}$ and $-1$ is assigned to the $\ket{0}$ state along $\hat{n}$. 
The main differences between $\hat{O}^{\rm{LP}}_{\alpha}$ and $\hat{O}^{\rm{SS}}_{\hat{n}}$ are that in the latter we assign a non-zero value to longitudinally polarized particles and that we consider spin-squared operators along any axis, not just in the transverse plane.
Using these measurements, we construct two Bell operators
\begin{align}
  & \hat{\mathcal{B}}^{\rm{LP}} = \hat{O}^{\rm{LP}}_{\alpha} \otimes (\hat{O}^{\rm{LP}}_{\beta} - \hat{O}^{\rm{LP}}_{\beta'}) + \hat{O}^{\rm{LP}}_{\alpha'} \otimes (\hat{O}^{\rm{LP}}_{\beta} + \hat{O}^{\rm{LP}}_{\beta'}),\label{LPop}\\
  & \hat{\mathcal{B}}^{\rm{SS}} = \hat{O}^{\rm{SS}}_{\hat{n}} \otimes (\hat{O}^{\rm{SS}}_{\hat{m}} - \hat{O}^{\rm{SS}}_{\hat{m}'}) + \hat{O}^{\rm{SS}}_{\hat{n}'} \otimes (\hat{O}^{\rm{SS}}_{\hat{m}} + \hat{O}^{\rm{SS}}_{\hat{m}'}). \label{SSop}
\end{align}
We can then calculate Bell observables and inequalities using absolute values of expectations of these operators
\begin{align}
  & \mathcal{I}_2^{\rm{LP}} \equiv  \left|\text{tr}\left(\hat{\mathcal{B}}^{\text{LP}}\rho\right)\right| \leq 2, \label{LPmax}\\
  & \mathcal{I}_2^{\rm{SS}} \equiv  \left|\text{tr}\left(\hat{\mathcal{B}}^{\text{SS}}\rho\right)\right| \leq 2. \label{SSmax}
\end{align}
\subsection{Choice of measurement settings}
Our goal is to observe states for which the value of a Bell observable exceeds 2. Therefore, we aim to find measurement settings (angles $\alpha, \alpha', \beta, \beta'$ for the LP-Bell operator and unit vectors $\hat{n}, \hat{n}', \hat{m}, \hat{m}'$ for the SS-Bell operator) that, for a given state, maximize its value.

The maximization of the LP-Bell expectation value over measurement settings $\alpha, \alpha', \beta, \beta'$ can be performed analytically using methods derived in~\cite{HORODECKI}. We obtain
\begin{equation}
  \label{lpbellsimple}
  \max\limits_{\alpha, \alpha', \beta, \beta'}\left(\mathcal{I}_2^{\text{LP}} \right)= 2\sqrt{\text{tr}\left(K^2\right)},
\end{equation}
where
\begin{equation}
K = \begin{pmatrix} \text{tr}\left(\lambda_{4}\otimes\lambda_{4} \rho\right) & \text{tr}\left(\lambda_{4}\otimes\lambda_{5} \rho\right) \\
  \text{tr}\left(\lambda_{5}\otimes\lambda_{4} \rho\right) & \text{tr}\left(\lambda_{5}\otimes\lambda_{5} \rho\right) \end{pmatrix}. \label{KMatrix}
\end{equation}

Maximizing $\mathcal{I}^{SS}_2$ across measurement directions $\hat{n}$, $\hat{n}'$, $\hat{m}$, and $\hat{m}'$ necessitated numerical methods in all nontrivial cases considered.
To enhance convergence, we do not optimize across all eight angles that define the unit vectors. Rather, we restrict them to be parameterized by five variables, $\alpha, \alpha', \beta, \beta', \theta$, such that
\begin{align}
  \label{eq:parm_1} 
  & \hat{n} = \hat{R}_{\theta} (\cos(\alpha), \sin(\alpha), 0)^T, \,\, \hat{n}' = \hat{R}_{\theta} (\cos(\alpha'), \sin(\alpha'), 0)^T, \\ \nonumber
  & \hat{m} = \hat{R}_{\theta} (\cos(\beta), \sin(\beta), 0)^T, \,\, \hat{m}' = \hat{R}_{\theta} (\cos(\beta'), \sin(\beta'), 0)^T,\\ \nonumber
  &\hat{R}_{\theta} = \begin{pmatrix} \cos(\theta) & 0 & -\sin(\theta) \\
                                          0       & 1 &      0        \\
                                     \sin(\theta) & 0 & \cos(\theta) \end{pmatrix}.
\end{align}
We also consider a simpler but complementary maximization over two variables, $\gamma$ and $\delta$, such that
\begin{align}
  \label{eq:parm2}
  & \hat{n} = (\sin(\gamma), 0, \cos(\gamma)), \,\, \hat{n}' = (\sin(\delta), 0, \cos(\delta)), \\ \nonumber
  & \hat{m} = \hat{n}', \,\, \hat{m}' = \hat{n}. 
\end{align}
The expectation value $\mathcal{I}^{SS}_2$ is calculated as the highest output from these two constrained maximization procedures. We find that this strategy produces satisfactory results. We leave the task of maximizing $\mathcal{I}^{SS}_2$ across all potential measurement settings for future work.

\subsection{Quantum state tomography}
In the absence of fiducial cuts on final-state particles, the reconstruction of the density matrix from angular distributions of particle velocities can be done using methods presented in~\cite{Ashby_Pickering_2023}. In this section, we present the results of applying this formalism to the LP-Bell and SS-Bell operators.

In the rest frame of each of the bosons considered, denoted 1 and 2, we define a spherical polar coordinate system such that $\theta_i$ is the polar angle measured from the axis of motion of particle $i$ and $\phi_i$ is the corresponding azimuthal angle, measured from the plane containing the direction of the beam that particle $i$ is moving closest to in this reference frame. 

In the case of gluons and photons splitting into fermions, one cannot generally neglect the fermion rest masses compared to the invariant mass of the off-shell gauge boson. To study the effect of massive final-state particles, we perform a tree-level calculation of the probability of a polarized off-shell gluon or photon with invariant mass $M_i$ splitting into two fermions with rest masses $m_{1i}$ and $m_{2i}$ along a given direction in the boson's rest frame. We find
that the modulation of this direction's angular distribution due to the mother particle's spin is smaller relative to that obtained in the $m_{1i}, m_{2i} \ll M_i$ limit by a factor
\begin{equation}
  \zeta_i \equiv \frac{2M_i^2 + (m_{1i} + m_{2i})^2}{2\left(M_i^2 - (m_{1i} + m_{2i})^2\right)}.
\end{equation}
For $W$ and $Z$ bosons, the rest masses of the daughter fermions can typically be neglected, so $\zeta_i \approx 1$. In general, the factor $\zeta_i$ must be included for each boson when averaging the Wigner $P$ symbols~\cite{Ashby_Pickering_2023} over events to extract elements of the density matrix.
Note that $\zeta_i$ diverges when the splitting is at the kinematic threshold, $M_i = m_{1i} + m_{2i}$. This results in a large statistical error when averaging over events. To mitigate this, a cut on the invariant mass of the fermion pair will be required in practice.

The remaining calculation follows~\cite{Ashby_Pickering_2023}. We obtain
\begin{align}
&\text{tr}\left(\hat{O}^{\rm{LP}}_{\alpha} \otimes \hat{O}^{\rm{LP}}_{\beta} \rho \right) \nonumber \\ & \quad = \left\langle 16 \zeta_1 \zeta_2 \cos(2(\phi_1 - \alpha))\cos(2(\phi_2 - \beta))\right\rangle, \label{LPBellTomo} \\
&\text{tr}\left(\hat{O}^{\rm{SS}}_{\hat{n}} \otimes \hat{O}^{\rm{SS}}_{\hat{m}} \rho \right) \nonumber \\ & \quad = \left\langle \left(\frac1{3} + 20 \zeta_1 P_2(\hat{n} \cdot \hat{d}_1)\right)\left(\frac1{3} + 20 \zeta_2 P_2(\hat{m} \cdot \hat{d}_2)\right)\right\rangle, \label{SSBellTomo}
\end{align}
where $\hat{d}_i = (\sin(\theta_i) \cos(\phi_i), \sin(\theta_i) \sin(\phi_i), \cos(\theta_i))$, $P_2(x) = 1/2 (x^2 - 1/3)$ is the second Legendre polynomial, and $\langle \cdot \rangle$ denotes the average over events.
The formulas for expectation values $\mathcal{I}_2^{\rm{LP}}$ and $\mathcal{I}_2^{\rm{SS}}$ are obtained by writing down linear combinations of the above formulas as specified by equations~\eqref{LPop} and~\eqref{SSop}.

Note that from equation~\eqref{SSBellTomo}, it follows that the SS-Bell observable is invariant under changes of basis, as long as one adjusts the settings $\hat{n}$ and $\hat{m}$ accordingly.
Thus, by maximizing the SS-Bell expectation value, we are also maximizing over the basis of the reconstructed fictitious state, similarly to the procedure performed for $t\bar{t}$ states in~\cite{fictitious_states}.

\section{Study of the inequalities for $ZZ$ and $W g(\rightarrow b \bar{b})$ states at leading order}
\label{study}
\begin{figure*}[t]
  \centering
  \includegraphics[width=0.98\textwidth]{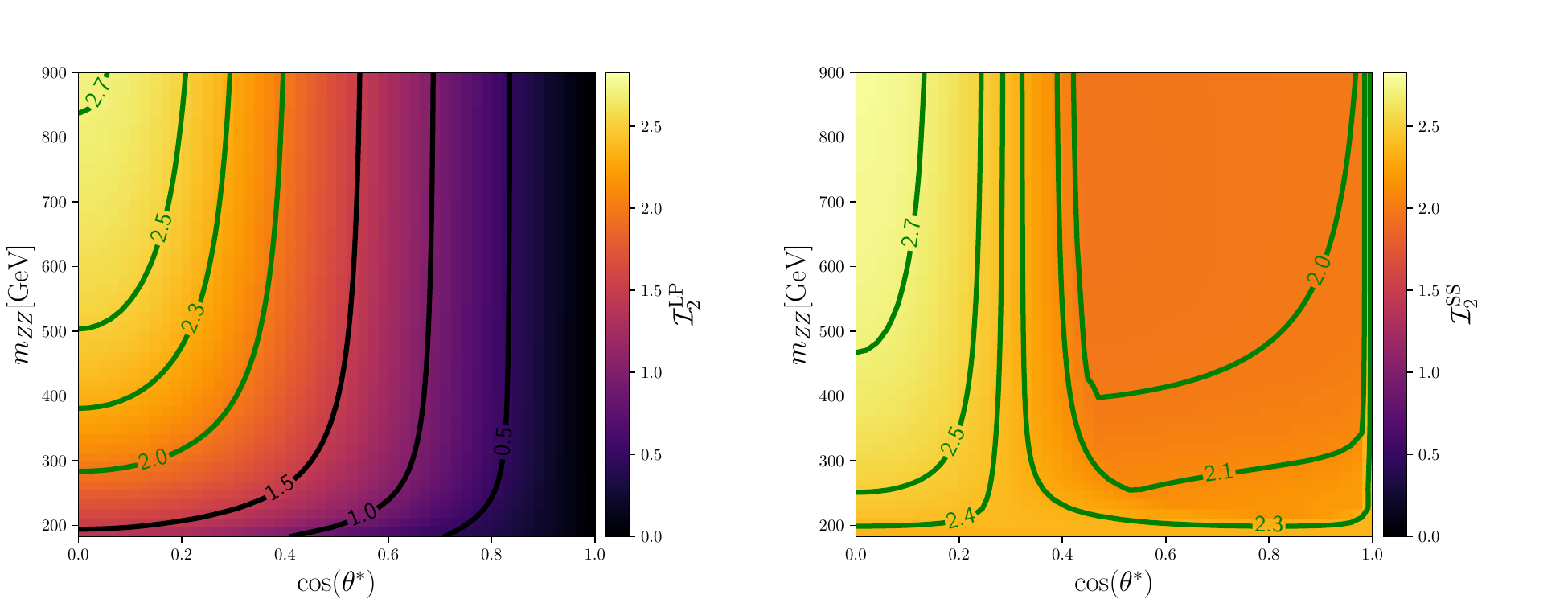} 
  \caption{Expectation value of the LP-Bell operator (left) and the SS-Bell operator (right) for $ZZ$ states obtained from quark fusion at leading order as a function of the invariant mass of the diboson system $m_{ZZ}$ and the smaller angle between $ZZ$ and the beam momenta in the partonic center-of-mass frame. The distribution is symmetric under $\cos(\theta^{*}) \rightarrow -\cos(\theta^{*})$. Bell violation is predicted for values larger than 2.}
  \label{fig:ZZcam} 
\end{figure*}

In $2 \rightarrow 2$ scattering, there is a symmetry under reflections about the event plane, corresponding to $y \rightarrow -y$ in our adopted basis. This implies
$\text{tr}\left(\lambda_{4}\otimes\lambda_{5} \rho\right) = \text{tr}\left(\lambda_{5}\otimes\lambda_{4} \rho\right) = 0$, simplifying the expression for $\mathcal{I}^{\text{LP}}_2$. We investigate its value analytically and compare it with the numerical results for $\mathcal{I}^{\text{SS}}_2$.
\subsection{$q\bar{q} \rightarrow ZZ$}
\begin{figure*}[t]
  \centering
  \includegraphics[width=0.98\textwidth]{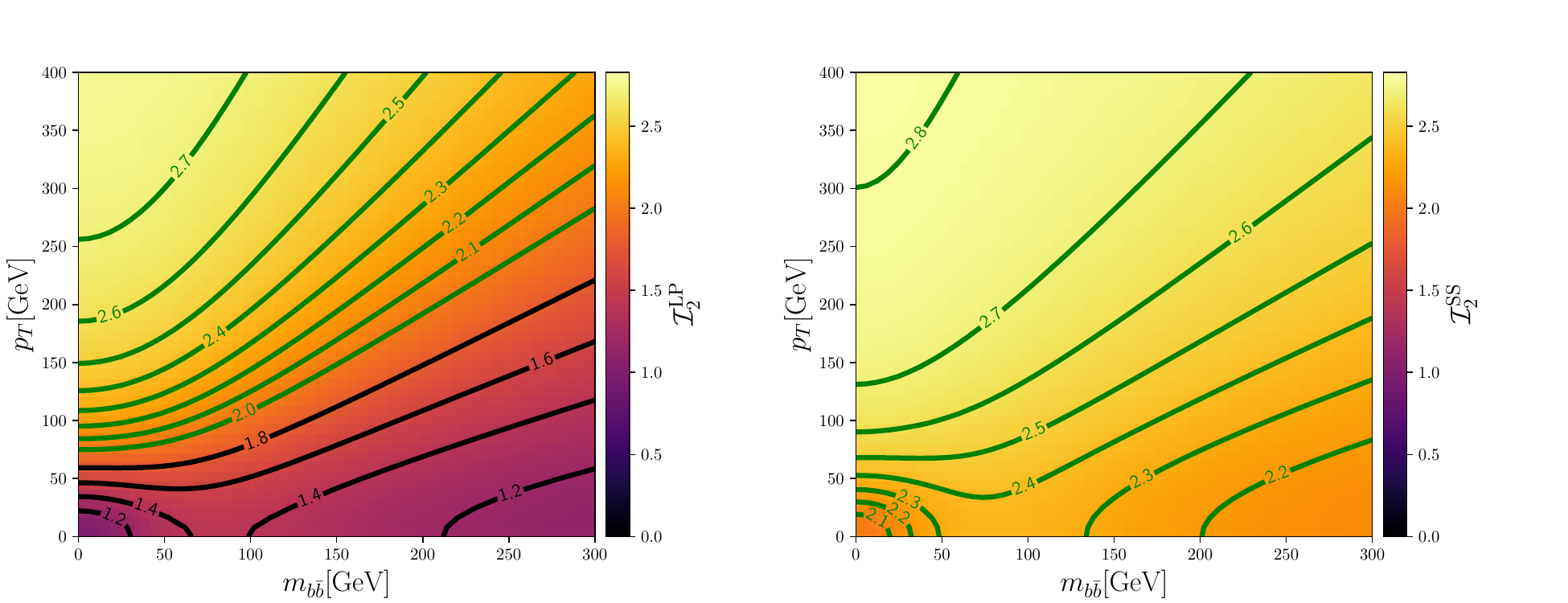} 
  \caption{Expectation value of the LP-Bell operator (left) and the SS-Bell operator (right) for $Wg$ states in the $\theta^{*} = \pi/2$ configuration at leading order as a function of the transverse momentum of the $b\bar{b}$ system relative to the beam $p_T$, and its invariant mass $m_{b\bar{b}}$. Bell violation is predicted for values larger than 2.} 
  \label{Wbbcentral} 
\end{figure*}
The results for this process are presented as functions of the center-of-mass energy $m_{ZZ}$ and the center-of-mass scattering angle to the beam $\theta^{*}$. The initiating quarks are taken to be massless. 
We evaluate $\mathcal{I}_2^{\text{LP}}$ analytically and $\mathcal{I}_2^{\text{SS}}$ numerically using amplitudes derived in~\cite{Aoude} and aforementioned measurement settings. The results are shown in figure~\ref{fig:ZZcam}. We discuss them by considering the low and high-energy limits.

Approaching the kinematic threshold, $m_{ZZ} \rightarrow 2 m_Z$, the density matrix becomes a mixture of two pure states related by a parity transformation
\begin{align}
  & \rho = p\ket{\Psi_{+}} \bra{\Psi_{+}} + (1 - p) \ket{\Psi_{-}} \bra{\Psi_{-}},
\end{align}
where $p$ depends on the initiating quarks’ left- and right-handed couplings to the $Z$ boson. Due to the parity invariance of operators considered, $\mathcal{I}_2^{\text{LP}}$ and $\mathcal{I}_2^{\text{SS}}$ are independent of $p$. The pure states are
\begin{align}
  \ket{\Psi_{\pm}} = \frac1{\sqrt{2}} \left(\ket{0}_{b} \ket{\pm}_b + \ket{\pm}_b \ket{0}_b\right),
\end{align}
where $\ket{\cdot}_b$ denotes a spin state along the beam direction.

The LP-Bell operator expectation value for this state has an ``unphysical'' dependence on $\theta^{*}$ stemming from the choice of basis included in the operator's definition.
We find that its maximum is at $\theta^{*} = \pi/2$ and is equal to $\sqrt{2}$, so the LP-Bell inequality is not violated in this limit.

The SS-Bell operator expectation value for this state is $\approx 2.36$ for all $\theta^{*}$. This value is reached with measurement settings defined by parameterization~\eqref{eq:parm_1} with $\theta = \theta^{*}$, corresponding to the ``beam basis''.
Our choice of operators allows Bell violation for states with low values of $m_{ZZ}$. This is particularly useful because the cross section is maximal in that kinematic regime.

In the high-energy limit, $m_{ZZ} \gg m_Z$, the $K$ matrix defined in equation~\eqref{KMatrix} becomes

\begin{align}
&K = \frac{\sin^2(\theta^*)}{1 + \cos^2(\theta^*)}\text{diag}(-1, 1).
\end{align}
We have
\begin{align}
  &\mathcal{I}_2^{\text{LP}} = \frac{\sin^2(\theta^{*})}{1 + \cos^2(\theta^{*})} 2 \sqrt{2},\label{I23MAXHE}
\end{align}
resulting in Bell violation when $\cos(\theta^{*}) < \sqrt{2} - 1 \approx 0.41$ and maximal violation when $\theta^{*} = \pi/2$.
The $\mathcal{I}_2^{\text{SS}}$ expectation value behaves identically for $\cos(\theta^{*}) < \sqrt{2} - 1$ in this limit, converging to the asymptotic result more quickly than the LP-Bell observable. It does not exceed 2 in the high energy limit when $\cos(\theta^{*}) > \sqrt{2} - 1$,
except for $\cos(\theta^{*}) \rightarrow 1$, where the spin state is identical to the one near threshold for all values of $m_{ZZ}$.\footnote{The discrepancy between our result and~\cite{Aoude} in terms of the spin state of a $ZZ$ system collinear to the beam arises due a to different ordering of the $m_{ZZ} \rightarrow +\infty$ and $\cos(\theta^{*}) \rightarrow 1$ limits in that work (first high-energy, second collinear), leading to a factorized spin state.}

While we have performed calculations for the $pp$ initial state, we observe that the findings of this section apply identically to the $ZZ$ final state at $e^{+}e^{-}$ and $\mu^{+}\mu^{-}$ colliders due to the cancellation of couplings of the initiating channel to the $Z$ bosons in expressions for the Bell operator expectation values.

To obtain an initial estimate of the SS-Bell observable at LHC, we simulate the $pp \rightarrow ZZ \rightarrow \text{leptons}$ process at leading order using $\texttt{MadGraph}$~\cite{MadGraph} and $\texttt{MadSpin}$ \cite{madspin}. 
We consider the 4-lepton final states and apply fiducial cuts on leptons identical to~\cite{ZZATLAS} i.e. $p_T > 7 (5)$ GeV and $|\eta| < 2.47 (2.7)$ cuts on electrons (muons). The leading and sub-leading leptons 
have an additional $p_T > 20$ GeV cut. The invariant masses of pairs of electrons and muons are required to satisfy $|m_{ll} - m_Z| < 10$ GeV, and the invariant mass of all four leptons is required to satisfy $m_{4l} > 180$ GeV.
We do not apply any further cuts that could enhance Bell violation in order to keep the statistical error minimal. The total number of events is estimated using the event yield presented in~\cite{ZZATLAS}, $\approx 3100$ events for the integrated luminosity of 140 fb$^{-1}$.
In the simulations, as well as in the estimated number of events, we consider final states with electrons or muons; however, in principle, any decay of $Z$ bosons, including semi-hadronic and fully hadronic, can be used.

In the presence of fiducial cuts, we cannot use equation~\eqref{SSBellTomo} to calculate $\mathcal{I}_{2}^{\text{SS}} $. Instead, it is estimated using the kinematic approach of~\cite{WithoutDecays}. 
From each simulated event, we extract $m_{ZZ}$ and $\theta^{*}$ and use the amplitudes from~\cite{Aoude} to calculate the density matrix and find the optimal measurement settings. We then use this density matrix and optimal settings to calculate $\mathcal{I}^{\text{SS}}_2$ for this event.
This approach was validated on a sample without fiducial cuts on final-state leptons, where we can use this method in parallel with equation~\eqref{SSBellTomo} and compare the results.
We find that the two methods are equivalent, allowing for statistical fluctuations. We use the standard deviation of the distribution of results from 100 independent repetitions of this test on a given expected number of events as an estimate of the statistical error on the observable.

We consider the Run 2 + Run 3 integrated luminosity of 300 fb$^{-1}$ ($\approx$ 6500 events) as well as the HL-LHC integrated luminosity of 3000 fb$^{-1}$ ($\approx$ 65000 events). The result is
\begin{equation}
\mathcal{I}_{2}^{\text{SS}} = 2.21 \,\,(\pm 0.19_{\text{stat., Run 2 + 3}})\,\,(\pm 0.06_{\text{stat., HL-LHC}}),
\end{equation}
corresponding to a 1.1$\sigma$ and a 3.5$\sigma$ significance of observing a Bell-violating state, respectively. 

\subsection{$q\bar{q} \rightarrow W (g \rightarrow b \bar{b})$}
Next, we turn to the production of a $W$ boson together with an off-shell gluon.
Unlike $W$ and $Z$ bosons, the gluon lacks a Goldstone mode, so regardless of its virtuality, it is in a spin-1 state. It must be off-shell to split into a massive $b\bar{b}$ pair, thus it has both transverse and longitudinal degrees of freedom.
We conclude that, like on-shell electroweak bosons, it can be treated as a qutrit.
The main difference compared to the on-shell $ZZ$ case is that the masses of bosons are now different, and the invariant mass of the virtual gluon state, $m_{b\bar{b}}$, is variable across events.

We initially consider two kinematic configurations, $\theta^{*} = \pi/2$, keeping $m_{b\bar{b}}$ and $p_T$ as variables, as well as $m_{b\bar{b}} \ll \{m_W, p_T\}$, keeping $p_T$ and $\theta^{*}$ as variables. 
The state is no longer symmetric under the exchange of bosons, so we assign $\hat{n},\, \hat{n}',\, \alpha, \,\alpha'$ to the gluon and $\hat{m}, \,\hat{m}',\, \beta,\, \beta'$ to the $W$ boson.

When $\theta^{*} = \pi/2$, we get
\begin{align}
  & K = \text{diag}
    \left(-\frac{E^2_g + E^2_W}{2 H^2} ,\frac{E_g E_W}{H^2}\right),
\end{align}
where $H^2 = m_{b\bar{b}}^2 + m_W^2 + p_T^2$ and $E_g,\, E_W$ are energies of the gluon and $W$ boson in the partonic center-of-mass frame of the collision. A direct calculation shows that in this configuration,
\begin{equation}
\mathcal{I}_2^{LP} > 2 \Leftrightarrow p_T > \left(\frac{3m_{b\bar{b}}^4 + 2 m_{b\bar{b}}^2 m_W^2 + 3 m_W^4}{4}\right)^{\frac{1}{4}}.
\end{equation}
On the other hand, numerical maximization of $\mathcal{I}^{\text{SS}}_2$ over measurement settings in this case shows that it is greater or equal to 2 for any value of $m_{b\bar{b}}$ and $p_T$, with equality only when $m_{b\bar{b}} \rightarrow 0$ GeV and $p_T \rightarrow 0$ GeV (we do not observe a dependence on the order in which the limits are taken). The quantum state factorizes in this limit, so there cannot be a Bell inequality violation.
We present the results of these calculations in figure~\ref{Wbbcentral}.

We now consider the limit $m_{b\bar{b}} \ll \{m_W, p_T\}$ for a general $\theta^{*}$. We show numerical results for $p_T > 5$ GeV to avoid instabilities arising from divergences in the amplitudes. 
The $K$ matrix becomes
\begin{align}
  & K = \text{diag}
    \left(-\frac{E^2_g + E^2_W}{2 H_{\theta}^2} ,\frac{E_g E_W}{H_{\theta}^2}\right),
\end{align}
where
\begin{equation}
H_{\theta}^2 = \frac{E_g^2(1 + \cos^2(\theta^*)) + m_W^2} {\sin^2(\theta^*)}.
\end{equation}
We conclude that the LP-Bell inequality is violated in this limit when
\begin{align}
\frac{p_T}{m_W} > s_{\theta} \sqrt{\frac{\frac{s_{\theta}^2}{2} \sqrt{(1-3c_{\theta}^2)(3-c_{\theta}^2)} + 2c_{\theta}^2 \left(3-c_{\theta}^2\right)}{\left((\sqrt{2}+1)^2 - c_{\theta}^2\right)\left((\sqrt{2}-1)^2 - c_{\theta}^2\right)}},
\end{align}
where $s_{\theta} = \sin(\theta^{*})$, $c_{\theta} = \cos(\theta^{*})$, and $|c_{\theta}| < \sqrt{2} - 1$. If $|c_{\theta}|$ is larger, there is no violation of the LP-Bell inequality.

\begin{figure*}[t]
  \centering
  \includegraphics[width=0.98\textwidth]{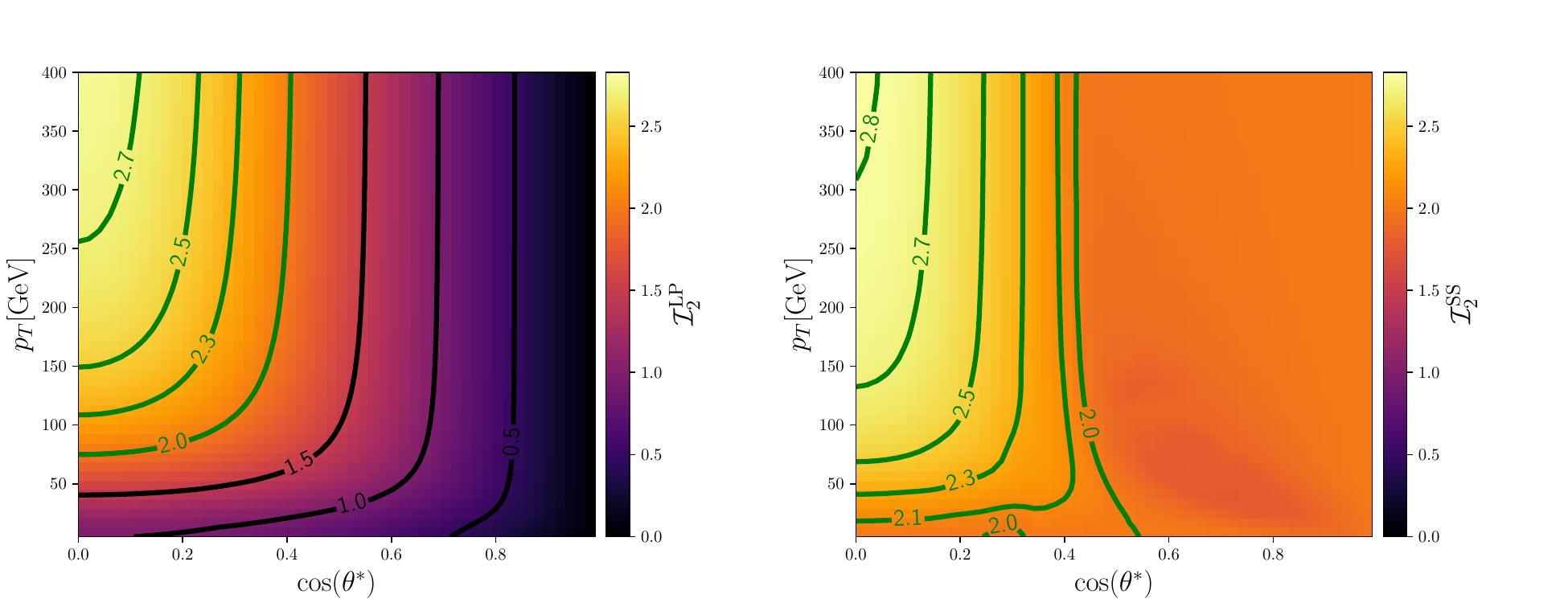} 
  \caption{Expectation value of the LP-Bell operator (left) and the SS-Bell operator (right) for $Wg$ states in the $m_{b\bar{b}} \ll \{p_T, m_W\}$ limit at leading order, as a function of the transverse momentum of the $b\bar{b}$ system, $p_T$, and the partonic center-of-mass scattering angle $\theta^{*}$ of the $Wg$ system. The distribution is symmetric under $\cos(\theta^{*}) \rightarrow -\cos(\theta^{*})$. Bell violation is predicted for values larger than 2.}
  \label{Wbbmgzero} 
\end{figure*}

For $\mathcal{I}^{\text{SS}}_2$, we show the results of numerical maximization for every point in the $\cos(\theta^{*})$-$p_T$ plane in figure~\ref{Wbbmgzero}. It appears that, at high-$p_T$, $|c_{\theta}| < \sqrt{2} - 1$ is a necessary condition for the violation of the SS-Bell inequality as well.
As expected, violation in this case can be seen at lower $p_T$ relative to the LP-Bell inequality.
We find that values of $m_{b\bar{b}}$ near $m_W$ generally give the largest values of the SS-Bell observable. Hence, the small $m_{b\bar{b}}$ limit serves as a lower bound on the achievable Bell violation.
When $m_{b\bar{b}} = m_W$, the results are identical to the $ZZ$ final state shown in figure~\ref{fig:ZZcam} up to the $W$ and $Z$ boson mass difference.

We obtain an estimate for $\mathcal{I}_2^{\text{SS}}$ at the LHC using the same methods as for the $ZZ$ final state.
With $\texttt{MadGraph}$, we simulate the $pp \rightarrow l \nu_l b \bar{b}$ process at leading order in all couplings. We apply similar fiducial cuts to~\cite{WBBCUTS}, i.e.
$p_{Tb} > 25$ GeV, $p_{Tl} > 30$ GeV, $|\eta_{b}| < 2.4$, $|\eta_{l}| < 2.1$. We do not include a cut on the angular distance between the $b$ and $\bar{b}$, 
as one could analyze events with two resolved $b$ jets and a single jet with a double $b$ tag.
We conclude from the studies presented in this section that an additional $|\cos(\theta^*)| < 0.3$ cut on the $Wg$ system is best to observe SS-Bell expectation values above 2. 
Statistical fluctuations in this case are very large without a cut on the mass of the $b\bar{b}$ system, because of the divergence of $\zeta_g$ when the quark-antiquark pair is produced at threshold. 
Hence, we also apply a $m_{b\bar{b}} > 20$ GeV cut.

We rely on the leading order cross section calculated by $\texttt{MadGraph}$ to estimate the number of events, noting that the $K$-factors can be controlled by exclusive cuts considered in~\cite{WBBCUTS}.
We also apply efficiency factors of 0.6 for observing two $b$ jets in this process and two factors of 0.7 for the identification of the lepton and missing energy. 
We get around 5000 events in Run 2 + 3 (300 fb$^{-1}$) and 50000 events in HL-LHC (3000 fb$^{-1}$), resulting in
\begin{equation}
  \mathcal{I}_{2}^{\text{SS}} = 2.35 \,\,(\pm 0.36_{\text{stat., Run 2 + 3}})\,\,(\pm 0.11_{\text{stat., HL-LHC}}),
\end{equation}
corresponding to a $< 1\sigma$ and a $3.1\sigma$ significance of observing a Bell-violating state, respectively. 
The increased statistical error relative to the $pp \rightarrow ZZ$ case is due to the large $\zeta_g$ factor of the gluon splitting to $b\bar{b}$. While more events
can be obtained by considering hadronic decay modes of the $W$ boson, we expect a large systematic error in the corresponding 4-jet measurement, as well as difficulties with separating background events.

We note that this process can also be studied using a variation of the CGLMP inequality from~\cite{Squared-spin-Bi}. However, since the SS-Bell inequality performs better than CGLMP in the $pp 
\rightarrow ZZ$ process with a similar topology, we chose not to consider it here.

\section{Resistance to noise}

In this section, we present the response of the Bell observables to a maximally mixed state, $\rho_{\text{mix}} = \mathds{1}_9/9$, which acts as a simplified model for irreducible backgrounds. For this state, the LP-Bell and SS-Bell observables give $\mathcal{I}_2^{\rm{LP}} = 0$ and $\mathcal{I}_2^{\rm{SS}} = 2/9$.
We take as an example a mix of this state with the $\theta^{*} = \pi/2$ high-energy limit $ZZ$ production at the LHC, for which the LP-Bell and SS-Bell operators have expectation values equal to $-2 \sqrt{2}$,
\begin{equation}
\rho = \lambda \rho_{ZZ} + (1 - \lambda) \rho_{\text{mix}},
\end{equation}
where $0 <\lambda < 1$. We obtain the following conditions on $\lambda$ for this state to be Bell-violating
\begin{align}
  & \lambda_{\text{LP}} > \frac{2}{2 \sqrt{2}} \approx 0.71,
  & \lambda_{\text{SS}} > \frac{20}{9 \cdot 2 \sqrt{2} + 2} \approx 0.73.
\end{align}
We conclude that the SS-Bell inequality exhibits a lower resistance to noise in this case. Note that for an expectation value of $+2\sqrt{2}$ the resistance to noise of the SS-Bell inequality would be higher; $\lambda_{\text{SS}}$ would be equal to $8/(9 \sqrt{2} - 1) \approx 0.682$, and $\lambda_{\text{LP}}$ would remain unchanged.

\section{Future perspectives}
In terms of phenomenology, it will be important to understand how higher-order EW and QCD corrections affect Bell observables defined here, similar to the study performed for other quantum observables in~\cite{vicini}. 

For the $W b\bar{b}$ process, $K$ factors, shown in~\cite{wbb1, wbb2}, would have to be controlled by exclusive cuts. The $K$ factors arise from a $gq$ initiating channel,
in which there is no intermediate gluon, so the density matrix we consider here would not be well defined.
Experimentally, the reconstruction of the neutrino momentum will be difficult due to the presence of two jets with uncertain
energies in the event. Another issue is a large $t\bar{t}$ background~\cite{CMSwbb}. Finally, some cuts may be required to control the contribution from double parton scattering \cite{DPS}.
Optimal ways of reconstructing the density matrix of the gluon in the presence of jets are also yet to be studied.

We expect other applications of the SS-Bell inequality to emerge in high-energy physics, given
the omnipresence of spin-1 particles produced by quark-antiquark fusion, for example, in meson decays.
Investigating the uses of the SS-Bell inequality in a broader quantum information context would also be interesting.

\section{Conclusions}

We discussed two Bell operators. The LP-Bell operator, considered before for scalar states of vector bosons in~\cite{Caban} was tested in diboson production via quark fusion, and the SS-Bell operator, which has not been used before in high-energy physics (and the author could not find a direct analog in other literature) was shown 
to perform exceptionally well in these circumstances.
Our new SS-Bell inequality greatly improves Bell violation prospects for states present in the $pp \rightarrow ZZ$ process compared to the CGLMP inequality. 

The inequalities discussed in this paper open Bell violation prospects to decays and splittings mediated by any interaction. Leveraging this, we presented the first prospect for measuring a Bell inequality violating state involving a gluon by considering the $(W \rightarrow l \nu)(g \rightarrow b\bar{b})$ process.

The studies contained in this paper provide a promising avenue for measuring Bell-violating states in $ZZ$ and $W + b\bar{b}$ events at the HL-LHC, assuming that one can control errors associated with the reconstruction of the gauge bosons involved.

\section*{Acknowledgements}
The author is grateful to Alan Barr, Fabrizio Caola, Claire Gwenlan, and Gavin Salam for their supervision, fruitful discussions, and comments on the manuscript.
The author would also like to thank the PanScales and the ATLAS $Z + b \bar{b}$ jet substructure analysis members for discussions during the early stages of this work.
The author is supported by an STFC studentship project ST/X508664/1, the Clarendon Scholarship, and the Wolfson Harrison UK Research Council Physics Scholarship.

\bibliographystyle{elsarticle-num}
\bibliography{biblio}
\end{document}